\documentstyle[11pt,aaspp4]{article}

\def\mathfont#1{\ifmmode{#1}\else{$#1$}\fi} %for math font     
\def\lae{\mathrel{<\kern-1.0em\lower0.9ex\hbox{$\sim$}}}  
\def\gae{\mathrel{>\kern-1.0em\lower0.9ex\hbox{$\sim$}}}  
  
\def\qo{\ifmmode{q_o}\else{$q_0$}\fi}  
\def\ho{\ifmmode{H_o}\else{$H_0$}\fi}   
\def\hounit{\ifmmode{{\rm km\  s}^{-1}\ {\rm Mpc}^{-        
1}}\else{${\rm    km\ s}^{-1}\ {\rm Mpc}^{-1}$}\fi}  
\def\zf{\ifmmode{z_{GF}}\else{$z_{GF}$}\fi}  
\def\kms{\ifmmode{{\rm km\ s}^{-1}}\else{${\rm km\ s}^{-1}$}\fi} 

\def\ergsec{\mathfont{ {\rm ergs\ s}^{-1}}}

\def\msun{\ifmmode{\ {\rm M}_\odot}\else{$ {\rm M}_\odot$}\fi}  
\def\msunyr{\ifmmode{\msun \ {\rm yr}^{-1}}\else{$\msun \ {\rm 
yr}^{-1}$}\fi}

\begin{document}
Accepted for publication in the Astrophysical Journal\\

\title{A Richness Study of 14 Distant X-ray Clusters From the 160 Square Degree Survey}

\author{B.R. McNamara,\altaffilmark{1,2} A. Vikhlinin,\altaffilmark{2}
A. Hornstrup,\altaffilmark{3} H. Quintana,\altaffilmark{5}
K. Whitman, \altaffilmark{4}
W. Forman,\altaffilmark{2} and C. Jones\altaffilmark{2}}

\altaffiltext{1}{Department of Physics \& Astronomy, Ohio University,
Athens, OH 45701, U.S.A.}
\altaffiltext{2}{Harvard-Smithsonian Center for Astrophysics, 60 Garden
St. Cambridge, MA, USA}
\altaffiltext{3}{Danish Space Research Institute,
Juliane Maries Vej 30, 2100 Copenhagen O, Denmark}
\altaffiltext{4}{Department of Astronomy, Cornell University, Ithaca, NY, U.S.A}
\altaffiltext{5}{Dpto. de
Astronomia y Astrofisica, Pontificia Universidad Catolica, Casilla
104, 22 Santiago, Chile}
\authoremail{}

\begin{abstract}
We have measured the surface density of galaxies toward 14
X-ray-selected cluster candidates at redshifts $z\gae 0.46$,
and we show that they are associated with rich galaxy
concentrations.  These clusters, having X-ray luminosities 
between $L_{\rm x}(0.5-2)\sim 0.5-2.6\times 10^{44}~ \ergsec$, are among the most 
distant and luminous in our $160$ square degree $ROSAT$ PSPC cluster survey.   
We find that the clusters
range between  Abell richness classes $0-2$,
and have a most probable richness class of one. 
We compare the richness distribution
of our distant clusters
to those for three samples of nearby clusters with similar X-ray luminosities.  
We find that the nearby and distant samples have similar richness
distributions, which shows that clusters have apparently
not evolved substantially 
in richness since redshift $z =0.5$. There is, however, a marginal tendency
for the distant clusters to be slightly poorer than nearby clusters,
although deeper, multicolor data for a large  sample would be required to
confirm this trend.  We compare the distribution of
distant X-ray clusters in the $L_{\rm x}$--richness plane to the
distribution of optically-selected clusters from the Palomar
Distant Cluster Survey.
The optically-selected clusters appear overly rich
for their X-ray luminosities when compared to X-ray-selected clusters.  
Apparently, X-ray and optical surveys do not necessarily sample identical
mass concentrations at large redshifts.
This may indicate the existence of a population of optically rich clusters
with anomalously low X-ray emission.  More likely, however, it reflects 
the  tendency for optical surveys to select unvirialized 
mass concentrations, as might be expected when peering along 
large-scale filaments.

\end{abstract}


\section{Introduction}
In hierarchical cosmologies, structure forms on broad mass scales from
the growth of density perturbations in the early Universe. 
The rate at which structure grows is governed by
the cosmological parameters, particularly
the density parameter $\Omega_0$ (White \& Rees 1978; Kaiser 1986).  
Dark matter halos on cluster mass scales
grow slowly at redshifts below $z\simeq 1$ in low
density, hierarchical model Universes and flat, lambda-dominated Universes, 
while rapid and continuous
growth is expected between $z\sim 0.5$ and the present in a flat, $\Omega_0\simeq
1$ Universe with zero cosmological constant
(Richstone, Loeb, \& Turner 1992, Luppino \& Gioia 1995, Eke, Cole, \& Frenk 1996, Bahcall,
Fan, \& Cen 1997, Mathiesen \& Evrard 1998, Viana \& Liddle 1998).

To the extent that X-ray emission traces
deep cluster potential wells, the abundance of X-ray clusters
at redshifts $0.4 \lae z \lae 1$ should probe $\Omega_0$ directly and with
high sensitivity.  Accordingly, several X-ray-selected samples of distant clusters have
been drawn from the $ROSAT$ archive to investigate this 
issue (e.g. Collins et al. 1997, Jones et al. 1998, Rosati et al. 1998, Vikhlinin et al. 1998a, Henry et al. 2001, Gioia et al. 2001).
These surveys have yielded no evidence for evolution of
cluster abundances for clusters with X-ray luminosities below $L_{\rm x}(0.5-2) \lae 3\times
10^{44}~\ergsec$ and redshifts below $z\sim 0.4$.
However, there is a growing body of evidence
for a mild but significant 
decrease in the abundance of the most X-ray-luminous
distant clusters (Henry et al. 1992, Vikhlinin et al. 1998b).

The degree to which clusters have matured into galaxy-rich,
virialized structures as a function of look-back time (redshift)
likewise depends on the cosmological parameters, and
is therefore an important aspect of cluster evolution (Kaiser 1986;
Frenk et al. 1996). However, it is difficult to select samples of clusters
to study the systematic properties of cluster galaxy populations--the degree to which clusters have evolved into
rich concentrations of galaxies, for example--while avoiding fatal
selection biases.
Because the optical detectability of distant clusters is itself
coupled to their degree of galaxy concentration, optically-selected samples
are not ideally suited to such studies. 
On the other hand, X-ray-selected clusters are
largely unbiased with regard to the galaxy population, and are therefore 
suited to studies of cluster galaxy evolution.
Furthermore, the existence of reasonably well-defined
relationships between X-ray luminosity,
temperature, and mass (Evrard, Metzler, \& Navarro 1996, Mushotzky \& Scharf 1997, Markevitch 1998)
make it possible to use X-ray flux-limited samples of clusters 
in the nearby and distant Universe to determine the degree to
which the galaxy populations have settled into their
cluster potential wells as function of time and cluster mass.
One measure of this process is cluster richness (Yee and Lopez-Cruz 1999).

In this paper we examine the richness 
distribution (i.e. the net number of galaxies 
encircled by the Abell radius within two magnitudes of the
third brightest galaxy) for 14 of the 28 
most distant clusters found in our $ROSAT$ 160 square degree
cluster survey (Vikhlinin et al. 1998a).  
We compare our cluster richnesses to
those of nearby clusters with similar X-ray selection criteria, and
to those for distant, optically-selected clusters 
(e.g. Bower et al. 1994, Holden et al. 1997, Couch et al 1991,
Postman et al. 1996), and we discuss
our results in the context of cluster evolution.

\section{Observations}

The clusters were discovered using the X-ray selection function 
described in Vikhlinin et al. (1998a).
Of the two to three dozen cluster candidates beyond $z\gae 0.45$ 
discovered in our survey, the 
14 for which we have the best optical data are listed in Table 1.  
The spectroscopic redshifts
measured for the brightest one to three galaxies
nearest the X-ray centroid are taken from Mullis et al. (2001, in preparation).
% (shown to three significant figures);
%photometric redshifts were measured using the magnitude of the 
%brightest cluster
%galaxy (shown to two significant figures).  Photometric redshift errors
%are typically $\Delta z = \pm 0.07$ (c.f. Vikhlinin et al. 1998a);
%spectroscopic redshift errors are typically $\Delta z = \pm 0.002$.
We obtained R-band Harris filter images of candidate clusters with the 1.2 m telescope
of the F.L. Whipple Observatory, the 1 m Las Campanis telescope,
the Danish 1.5 m telescope, and the 3.6 m telescope at La Silla during many  observing runs
from June 1995 to February 1999. The optical images were
originally obtained to determine whether the extended X-ray sources
correspond to optical galaxy concentrations, and to measure
photometric redshifts for the clusters. Several CCD cameras were used over 
the course of the survey. Exposure times ranged from 10 min
to over an hour, and seeing was typically 1.3 arcsec.
The fields of view were 10.6 arcmin, 13.3
arcmin, 23 arcmin, and 5.5 arcmin for the 1.2 m, 1.5 m, 1 m, and 3.6 m telescopes
respectively.  For reference,  2.3 arcmin
corresponds to a linear size of 1 Mpc at $z=0.5$ for
${\rm H}_0=50~{\rm km}~{\rm s}^{-1}~{\rm Mpc}^{-1}$ and $q_0=0.5$,
which is assumed throughout this paper.  Deep images of Cl 0529-5848
and Cl 1311-0551 were taken with the EFOSC camera on the 
3.6 m telescope at La Silla. These images provided a deep measurement
of the galaxy luminosity function within a 1 Mpc radius for the
two clusters. We used the average of  these luminosity functions
to calculate magnitude depth incompleteness corrections
when estimating richnesses for the remaining clusters. 
All CCD image frames were de-biased, 
overclock-corrected, flat-fielded, and combined using IRAF routines to form the 
science images for each cluster.

%\section{Data Analysis}

\section{Galaxy Selection and Photometry}

The galaxies were selected for the analyses described below 
as follows.
Using a wavelet detection routine, we determined the locations of all 
sources on the science CCD frames.  
We measured the full width at half maximum (FWHM) in right ascension and
declination of each object, and rejected all point-like objects 
and bright foreground stars and galaxies.
We measured instrumental magnitudes
for each galaxy using 3 arcsec and 6 arcsec
diameter apertures, which correspond to 22 and 44 kpc
linear diameters at $z=0.5$. 

We then ranked the galaxies by magnitude relative to
the brightest galaxy nearest the X-ray 
centroid.  The brightest cluster galaxy and one or two companions
were used to determine the spectroscopic
redshift of the cluster (Mullis et al. 2001, in
preparation).  Galaxies within the Abell radius brighter than
$m_3 + 2$, where $m_3$ is the third brightest galaxy magnitude,
were selected from the catalog.  

We counted galaxies in the magnitude-selected
subsample of cluster and field galaxies within centered apertures
of $0.25, ~0.5, ~1.0,$ and $1.5$ Mpc radius.
We determined the background by averaging the counts in 20--30 apertures
placed at random locations away from the cluster.
The net number of cluster galaxies was estimated  by subtracting
the mean background counts, adjusted appropriately for
area, from the counts in the cluster aperture.

We restricted our search
for the third brightest cluster galaxy to a radius of 0.5 Mpc to minimize
foreground 
galaxy contamination that can introduce errors into
the richness measurements (Postman et al. 1996).  Using the integrated luminosity
function discussed in \S 6.1, we find that a 
half magnitude error in the counting depth can cause a $\approx 60\%$
error in the richness estimate. Errors of this magnitude can
be introduced when a bright foreground galaxy is misidentified as the
third brightest cluster galaxy (e.g. Postman et al. 1996). 
This would cause $m_3$ to be underestimated, which in turn would cause
us to mine to erroneously shallow depths.  This chain of errors
would then result in erroneously small
cluster richness estimates.  Applying the Abell radius
criterion to distant cluster fields would produce this error
most of the time (e.g. Postman et al. 1996). 
Therefore, we mitigated this effect by selecting
the third brightest galaxy within a half
Mpc radius.  

We determined whether we were successful at mitigating the third
brightest galaxy problem by comparing our distribution of 
$m_1-m_3$ to the distribution in the ACO catalog. 
We found good agreement between the two samples.
The mode for our 14 cluster sample
is $m_1-m_3\approx 0.4$ magnitudes, which compares well to
the ACO value of $\approx 0.35$ magnitudes. 
We therefore do not appear to be 
significantly biased in this respect.
Nevertheless, using another method, 
we were able to circumvent this process entirely
by comparing galaxy counts in nearby and distant clusters within 
restricted apertures and depths in \S 4 and \S 5.
The analysis was done using our own routines written in IDL and FORTRAN.
We compared our galaxy selection algorithm to the ``SExtractor''
routine and found similar results.  We show an optical image, with
X-ray contours superposed, of a
typical richness class 0 cluster discovered in our survey in Figure
1.  The brightest cluster galaxy nearest the X-ray centroid, 
indicated by an arrow, has a spectroscopic redshift of $z=0.531$.

\begin{figure}
\plotone{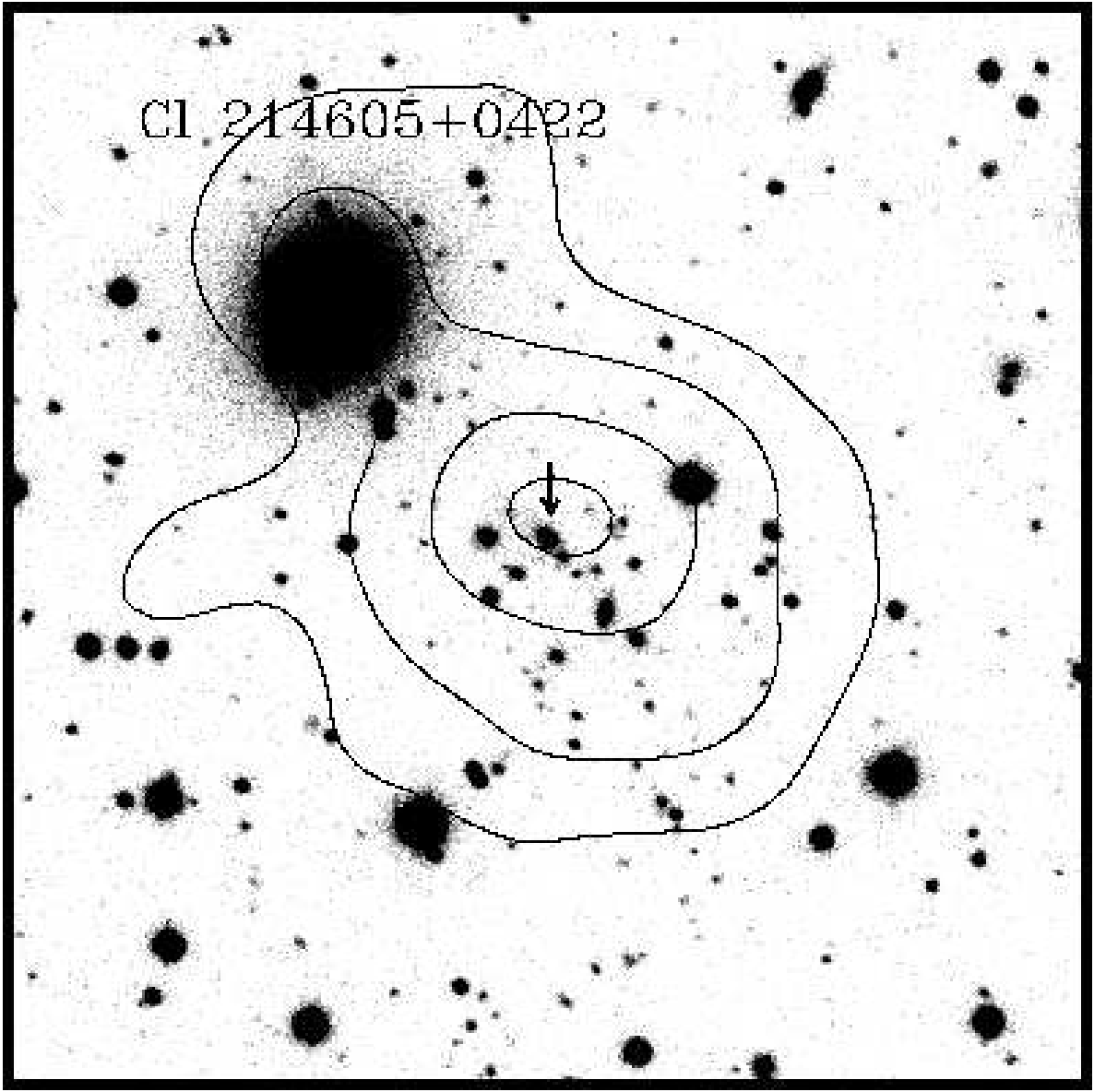}
\caption{ \label{fig1}A 35 minute exposure R-band image of the cluster
Cl214605+0422 (grayscale) obtained with the FLWO 48 inch telescope
with $ROSAT$ PSPC X-ray contours superposed.  The redshift
of the brightest cluster  galaxy (indicated by the arrow) was found to be
$z=0.531$, using spectra obtained with the MMT.  This object is
one of the distant clusters detected in the 160 square degree survey.  The
image is 4.4 arcmin on a side. North is at the top; east is to the left. 
}
\end{figure}

\section{The Cluster Contrast Parameter}

Because we were required to optically vet over 200  cluster candidates, we were
initially unable to obtain deep images for all clusters.  We therefore
obtained optical images to a sufficient depth to determine whether
the X-ray cluster candidates are indeed associated with galaxy concentrations,
and to measure brightest cluster galaxy magnitudes
for photometric redshift estimation (Vikhlinin et al. 1998 a, b).
In most instances, our data do not completely sample
cluster galaxies two magnitudes fainter than the third brightest galaxy.
Therefore, we were required to correct
our galaxy counts to a consistent depth, as discussed in \S 6.1.
However, first we will demonstrate that galaxy concentrations are
indeed present at the locations of the extended X-ray sources using
the contrast parameter introduced by Couch et al. (1991).

The contrast parameter
$\sigma_{\rm cl}=N_{\rm cl}/\sigma_{\rm bg}$ (Couch et al. 1991),
is a measure of the observed over-density of bright cluster galaxies
against the background.  Here, $N_{\rm cl}$ is the net number of galaxies 
within an aperture centered on the cluster, after subtracting 
the average background surface density measured in 20--30
randomly placed apertures surrounding the
cluster. The parameter $\sigma_{\rm bg}$ is the standard deviation about
the mean background value for the surrounding apertures.
The contrast parameter provides a
crude measure of cluster richness that is relatively insensitive 
to the imaging depth, but depends
strongly on redshift, central galaxy concentration,
and detection aperture size.

\begin{figure}
\plotone{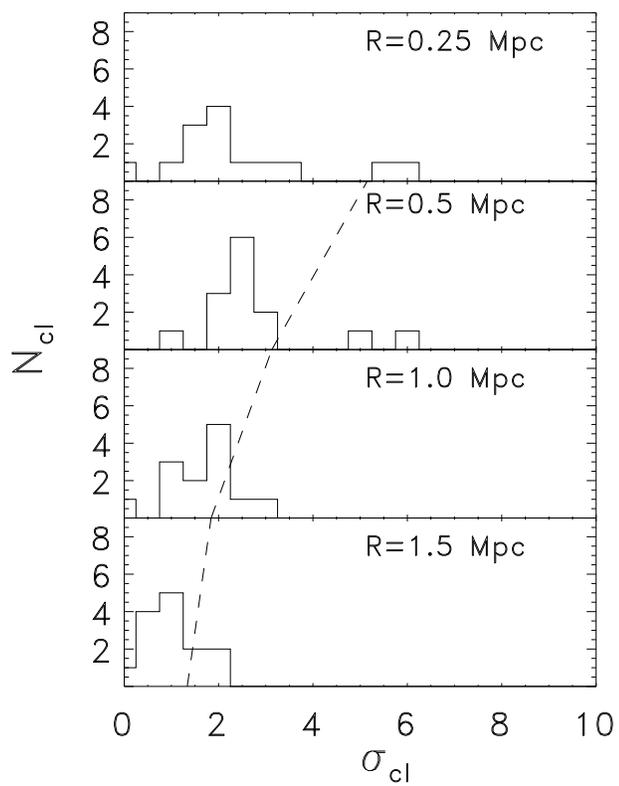}
\caption{Histogram of the contrast parameter $\sigma_{\rm cl}$ 
as a function of aperture size.  $\sigma_{\rm cl}$ 
decreases with increasing aperture size in a manner that is consistent
with an isothermal galaxy surface density profile (broken line). \label{fig1}}
\end{figure}

We demonstrate the effects of increasing aperture size and increasing
imaging depth on $\sigma_{\rm cl}$ in Figures 2 \& 3.
In Figure 2 we examine the effect of aperture size on the contrast 
parameter by plotting histograms of $\sigma_{\rm cl}$ against aperture
size for our sample.
$\sigma_{\rm cl}$ reaches a maximum
with the 0.25--0.5 Mpc apertures and declines with increasing aperture size.
Within these apertures, the modal contrast is $\simeq 2.5\pm 1$.
The contrast for an isothermal distribution of galaxies should
decline with aperture radius
roughly as $\sigma_{\rm cl}(r) \sim \ln r/r$ (shown as the broken line
in Figure 2), while $\sigma_{\rm cl}(r)\sim {\rm constant}$ for a
$\rho_{\rm g}\sim r^{-1}$ profile.  The contrast
parameter appears to decline with increasing aperture size in a
manner that is roughly consistent with an isothermal profile.
%Lubin \& Postman (1996) found on average a
%shallower, $r^{-1.4}$ surface density profile for their distant, 
%optical clusters, but our X-ray-selected clusters do not 
%seem to follow this profile.  

In Figure 3 we plot the contrast parameter within a 1 Mpc radius
as a function of imaging depth.  The imaging depth, $\Delta m$,
is the limiting magnitude depth relative to the brightest cluster galaxy's
magnitude.  The data are for two richness class 1--2 clusters in
our sample with deep imaging obtained with the ESO 3.6 m telescope.
No strong dependence is found between contrast significance
and imaging depth for either cluster.

Figures 2 \& 3 demonstrate that the optical contrast of clusters
against the background is relatively constant with imaging depth,
but declines rapidly with aperture size.  Both properties are due
to the rapidly increasing background galaxy counts.  
We find that a 1 Mpc aperture strikes a
good compromise between the competing demands of good
cluster counting statistics and minimizing the background noise.
In general, the background counts are
factors of 2--3 larger than the cluster counts within a 1 Mpc aperture.

Figures 2 \& 3 also show that relatively shallow imaging is effective
at detecting galaxy concentrations at redshift $z\sim 0.5$.
However, as is seen in Table 1 and discussed below, the contrast
parameter is not a reliable measure of cluster richness.
For instance, Couch et al. found that, within a 1.5 Mpc diameter
aperture using a passband similar to ours, $\sigma_{\rm cl}\sim 3$
corresponds approximately to a richness class 1 cluster, and
$\sigma_{\rm cl}\sim 4-6$ corresponds to a richness class 2 to 3
cluster at redshifts $z=0.5-0.7$.  We do not find such a large increase
in contrast with richness for our clusters (cf. Table 1).
On average, our clusters show a lower contrast than
Couch's, and fall in the regime where their catalog becomes 
seriously incomplete. In fact, a cluster such as Cl 2146+0422, shown
in Figure 1, would probably have been omitted from the
Couch et al. (1991) catalog.

\begin{figure}
\plotone{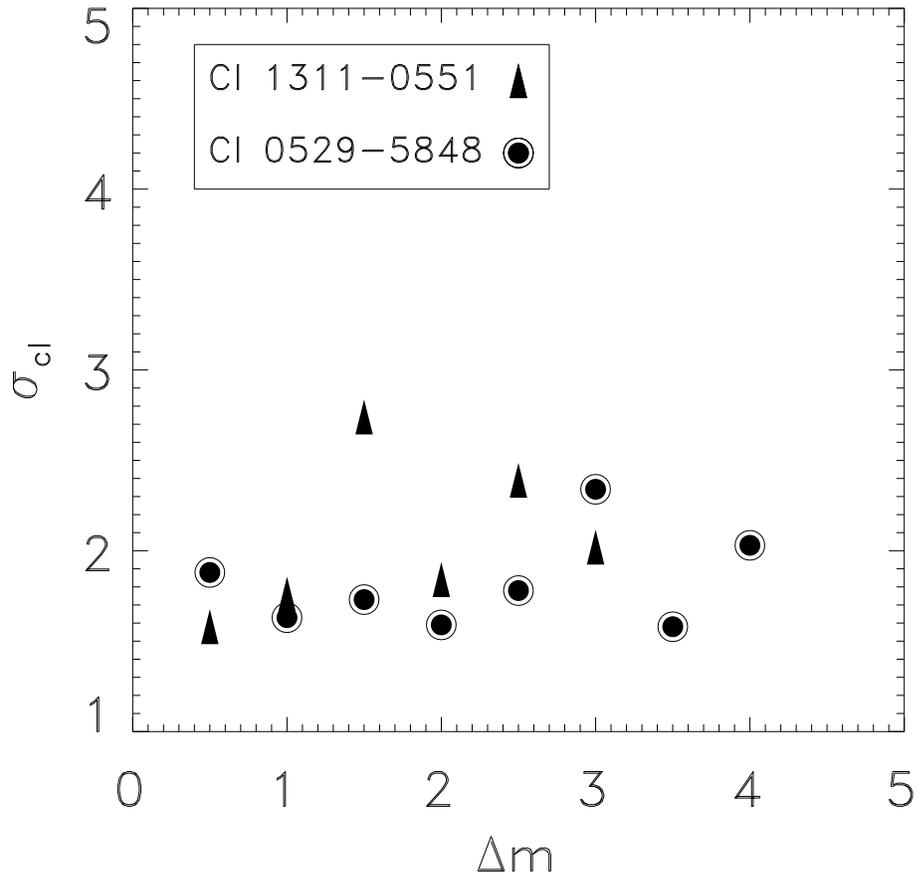}
\caption{Contrast parameter, $\sigma_{\rm cl}$, 
plotted against imaging depth, $\Delta m$,  for two deeply-imaged clusters.  $\sigma_{\rm cl}$ 
shows no significant increase with imaging depth. \label{fig1}}
\end{figure}

\section{Comparison of Galaxy Counts Between our Distant X-ray Clusters 
and Nearby Abell Clusters}

Here we compare the galaxy counts from our sample
to those in nearby clusters by comparing the net number
of galaxies within a 1 Mpc radius as a function of imaging depth.
This approach has two significant advantages over direct richness estimates.
First, galaxy counts  within a 1 Mpc radius are more accurate
than counts within a 3 Mpc radius, because the 
background corrections are smaller.  Second, we avoid entirely the uncertain 
depth and aperture corrections required to estimate Abell richnesses.
The primary disadvantage is that galaxy counts as a function of 
cluster radius and magnitude depth are generally unavailable for clusters.
We remedied this situation by measuring galaxy counts toward
14 nearby Abell clusters from Digital Sky Survey (DSS) images.
The clusters, with redshifts between $z=0.04-0.1$,  were selected from the 
ROSAT Brightest Cluster Sample (BCS, Ebeling et al. 1998). 
Their X-ray luminosities, $L_{\rm x}(0.1-2.4)=1-6\times 10^{44}~\ergsec$,
roughly match the range for our distant sample, after correcting for 
the bandpass offset. The galaxies on the DSS images were selected and
counted automatically, after correcting for photographic nonlinearity,
in a nearly identical manner to our distant cluster method.  
For example, the galaxy selection algorithm and background field
corrections were applied in a similar fashion.
The expression used
to compute the galaxy and background fluxes that relates intensity, $I$, to photographic density, $D$, is:
$\ln I=f_1\times D + f_2\times \log(e^{BD^{C_1}}-1) +f_3\times e^{BD^{C_2}} +f_4$.  The photographic nonlinearity was removed through
the coefficients $f$, $B$, and $C$, which were derived by comparing the
digitized photographic magnitudes of stars and galaxies from
the DSS to calibrated CCD images of the same fields.

We avoided the potential errors associated with selecting the
third brightest galaxy (cf. \S 3) by measuring depth with respect to the
magnitude of the BCG.
The largest uncertainty associated with this approach
would
be differential luminosity evolution of the BCG with respect
to the remaining cluster galaxies.  However, we don't believe
this to be a serious
problem.  The BCGs in our sample are
very good standard candles to $z=0.5-0.7$ (Vikhlinin et al. 1999).
The variation about their mean $R$-band luminosity is
$\sigma \approx 0.3$ magnitudes.  While a variation at this level
would affect individual clusters,
it would not significantly affect the comparison, as both samples
were treated in a similar fashion. Nevertheless,
this approach provides a good check on our application
of the Abell richness criteria.
The details of the automated DSS galaxy counts for the nearby
clusters will be presented in a future paper (Whitman, McNamara,
\& Vikhlinin, in prep).  It is worth noting in advance, however,
that our automated counts agree
with the ACO catalog to within the counting statistics for
all clusters within this luminosity and richness range.  
This check provides a measure of reassurance
that our comparison in Figure 4 for the 1 Mpc apertures is reliable.

\begin{figure}
\plotone{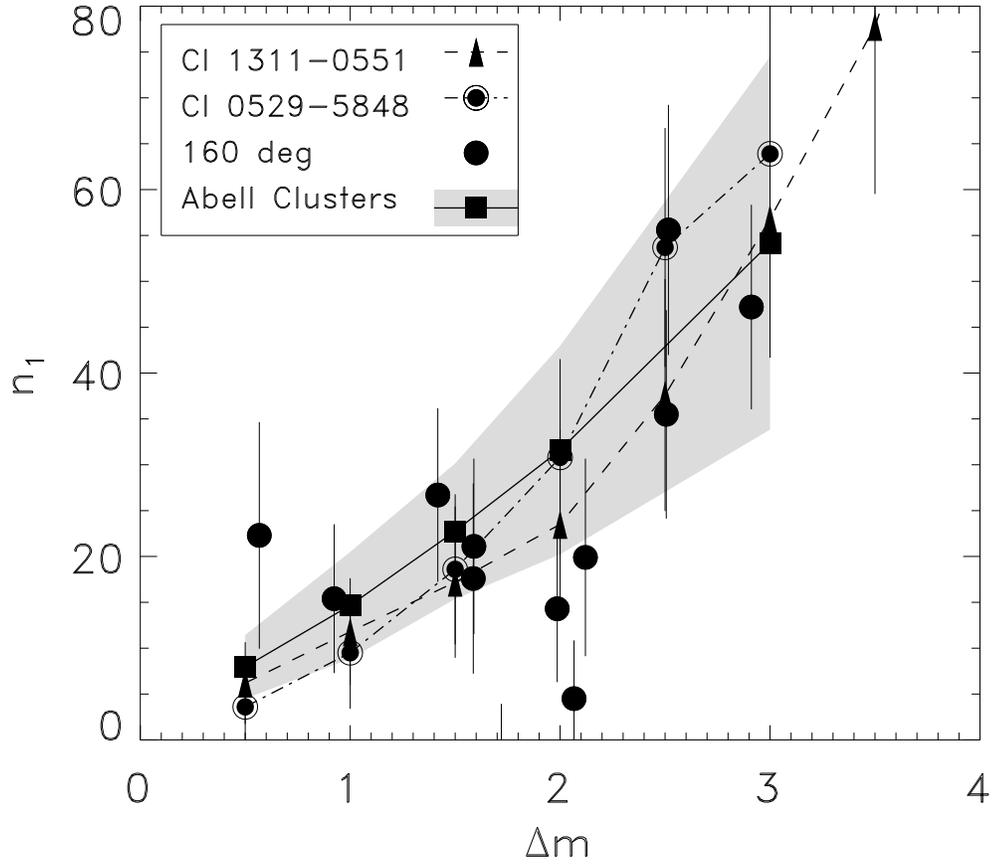}
\caption{ \label{fig1} Plot of galaxy counts within a 1 Mpc
radius aperture, $n_1$, vs the limiting imaging depth with
respect to the BCG, $\Delta m$.  The shaded region shows the
$1\sigma$ plane occupied by nearby, X-ray-selected Abell 
clusters.  The mean is indicated by connected, filled rectangles.
The 160 square degree clusters are shown as indicated in the
legend. }
\end{figure}

The galaxy count comparison between the BCS and 160 square degree
survey clusters as a function of imaging
depth is shown in Figure 4.  The average number of
galaxies as a function of depth within a 1 Mpc radius for
the 14 nearby BCS Abell clusters is shown as  connected, filled rectangles.
The one $\sigma$  deviation about the mean is 
enclosed by the shaded region.  The galaxy counts from the deep
imaging of Cl 1311-0551 and Cl 0529-5848 are shown as connected
points.  The remaining 160 square degree clusters are shown as solid
points.  The error bars are one sigma Poisson estimates using
the net cluster and background counts in each aperture,
calculated as $\sigma=\sqrt{N_{\rm cl}+N_{\rm bg}}$.

Figure 4 shows that nearly all
of the 160 square degree clusters lie within $2\sigma$ of the 
BCS clusters.  Although 
a few outliers are good candidate galaxy-poor clusters,
the overall agreement is quite good.  We can conclude
that the 160 degree survey clusters are similarly rich 
to the nearby BCS clusters with the same range of X-ray luminosity.
In the next section, we estimate the richnesses of the distant clusters
according to the Abell criteria 
by applying depth and aperture corrections to the galaxy counts.

\section{Estimating Cluster Richnesses}

\subsection{Imaging Depth Corrections}

Our images often do not reach a sufficient depth
to sample completely to $m_3+2$. 
Therefore, we measure an incompleteness parameter for each
cluster by taking the difference between
the location of
the peak of the instrumental magnitude histogram for all galaxies in 
each field and $m_3+2$. The incompleteness parameter, $\delta m$,
is listed in Table 1.  We corrected the galaxy counts assuming
a cumulative luminosity function of the form $\log N(\le m)=K +
S\times m$, where $S$ is the slope of the cumulative luminosity function
(Sarazin 1986).  The function was normalized by the detected number
of cluster galaxies,  $N_{\rm g}$, as
$\log N_{\rm g}^{\prime}=\log N_{\rm g} + S\delta m.$ The cluster
galaxy luminosity function is poorly known for distant clusters.  We
therefore estimated $S$ for
the sample using deep $R$-band images of Cl 1311-0551 and Cl 0529-5848,
which we obtained with the European
Southern Observatory's 3.6m telescope.
The cumulative distributions of galaxies, after background subtraction,
are shown for each cluster in Figure 5.
The curves shown are normalized to the magnitude of the brightest cluster
galaxy in each cluster. Our data sample completely three
magnitudes below the brightest galaxy for Cl1311-0551
and four magnitudes below the brightest galaxy for
Cl1311-0551. As the modal difference between the brightest and
third brightest galaxies for our 14 cluster sample
is $m_1-m_3\approx 0.4$ magnitudes (the ACO value is
$\approx 0.35$ magnitudes), these two clusters, if representative
of our sample as
a whole, reach a depth well below $m_3 + 2$. 
The slope of the linear fit to the data is 
$S=0.39\pm 0.1$.  We assume from now on that the luminosity functions 
of the remaining clusters have
this slope, and we use $S$ to correct their galaxy counts
for imaging depth incompleteness as described above.

\begin{figure}
\plotone{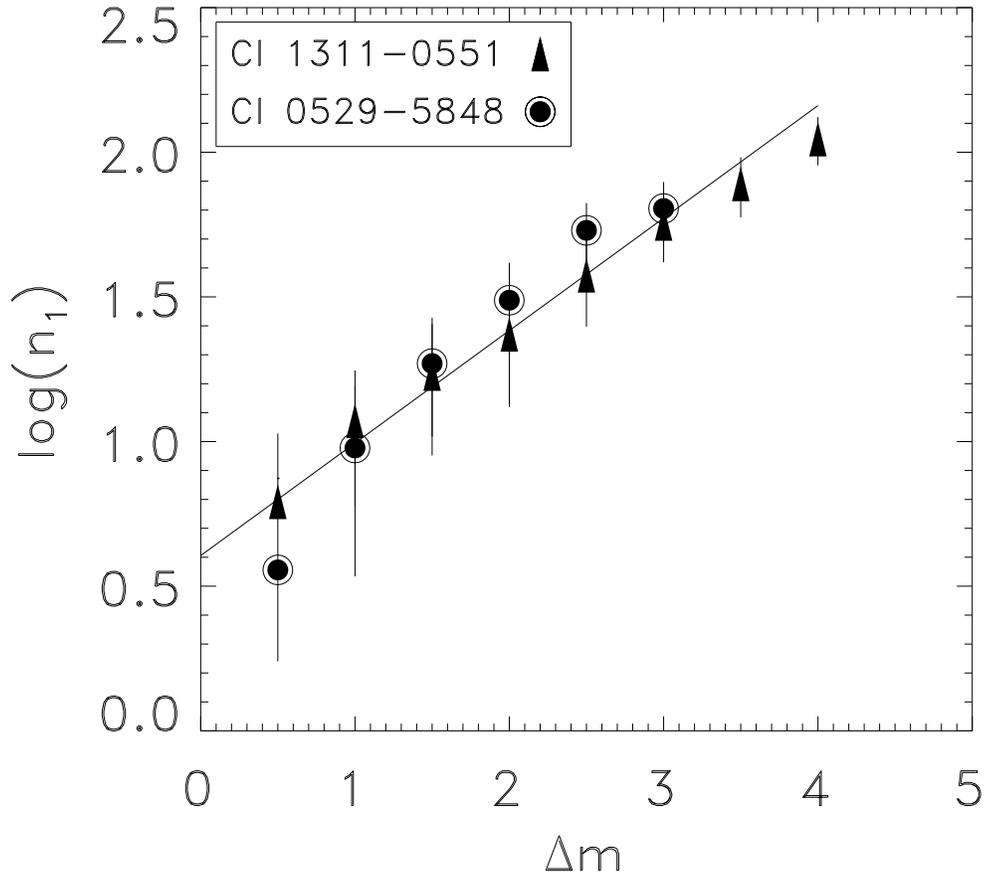}
\caption{ \label{fig1}Plot of the logarithm of galaxy counts within a 1 Mpc
radius aperture, $n_1$,  vs the limiting imaging depth with
respect to the BCG, $\Delta m$,  for two, deeply-imaged clusters.
The straight line fit to the data was used to apply the depth 
corrections to the remaining 160 degree survey objects. }
\end{figure}

\subsection{Comparison Between Nearby and Distant X-ray-Selected Cluster
Richnesses}

\vskip 20pt

After correcting our counts for imaging depth incompleteness,
we computed cluster richnesses, $\Lambda_{\rm c}$, 
by extrapolating counts in each aperture
to the 3 Mpc Abell radius.  The extrapolations assume both
$r^{-2}$ and $r^{-1.4}$ surface density 
profiles, normalized by the corrected counts within each aperture. 
Extrapolation, rather than direct measurement,
is required because the galaxy background 
increases more rapidly with increasing
aperture than the cluster counts.
When the size of the Poisson variations in the background counts approaches
the cluster signal, the cluster counts become
unreliable.  Although the backgrounds in the
0.25 and 0.5 Mpc apertures are only $\sim 50\%$ of the cluster counts, 
only a few cluster galaxies are generally present there.
Therefore, the Poisson errors in the cluster counts are relatively
large.  On the other hand, 
as the aperture grows to 1 Mpc,  the background becomes factors of
$\sim 2-3$ larger than the cluster counts.  Nonetheless, 
the one to three dozen net cluster galaxies present can be counted 
with greater precision.

Whether or not a universal galaxy surface density
profile for distant clusters exists is unknown.
The density profile of galaxies for a sample of 14
intermediate redshift ($z\sim 0.4$), X-ray-selected
clusters was found by Carlberg et al. (1997) to 
follow a Navarro, Frenk, \& White (1997)
profile.
This profile implies a  steep $r^{-3}$ decline in galaxy surface density
at very large radii, but a somewhat shallower decline
over much of the 3 Mpc region of interest.
On the other hand, the $z\gae 0.5$
clusters selected optically in the Palomar Distant Cluster Survey
appear to have considerably shallower, $r^{-1.4}$ profiles within
the radial range considered in this sample.

We therefore analyzed our sample by applying both $r^{-1.4}$
and isothermal, $r^{-2}$ surface density profiles to our aperture
corrections. In the upper panel of
Figure 6, we present richnesses for 13 clusters with $\delta m <1.5$. 
In the lower panel, we plot richnesses for a subset of 9 clusters with
smaller depth corrections, $\delta m < 0.7$.
The solid distribution of clusters has been
corrected assuming the $r^{-2}$  profile,
and the dashed distribution has been corrected for the $r^{-1.4}$
profile.  The shallower, $r^{-1.4}$ profile correction gives somewhat
higher richnesses compared to the $r^{-2}$
profile, as would be expected.  However, 
the distributions do not differ greatly.
Therefore, for simplicity, we adopt the $r^{-2}$ profile when deriving
richnesses and richness classes (e.g., Figure 7 and  Table 1),
although our conclusions do not depend significantly on this assumption.

The cluster richnesses and richness classes
were determined by applying the depth and aperture corrections to the 
$0.25, ~0.5$, and 1 Mpc aperture counts prior to averaging them.
Cluster richnesses are binned into richness classes ($RC$) mapped 
following the ACO catalog:
$\Lambda_{\rm c}=~10-30,~ 31-49,~ 50-79,~ 80-129,~130-200,~200-300$ 
$\rightarrow RC= -1,~0,~1,~2,~3,~4$, respectively.
The errors in the richnesses are dominated by systematics,
so we adopted the extreme values from the three apertures
as an error estimate (shown in Figure 7 and discussed below).
Figure 6 shows that our clusters are primarily richness class
0--2.  While three clusters may lie in richness classes $3-4$, 
their richness estimates depend on large and uncertain depth 
corrections (see Table 1).  Their richness estimates are therefore
suspect.

\begin{figure}
\plotone{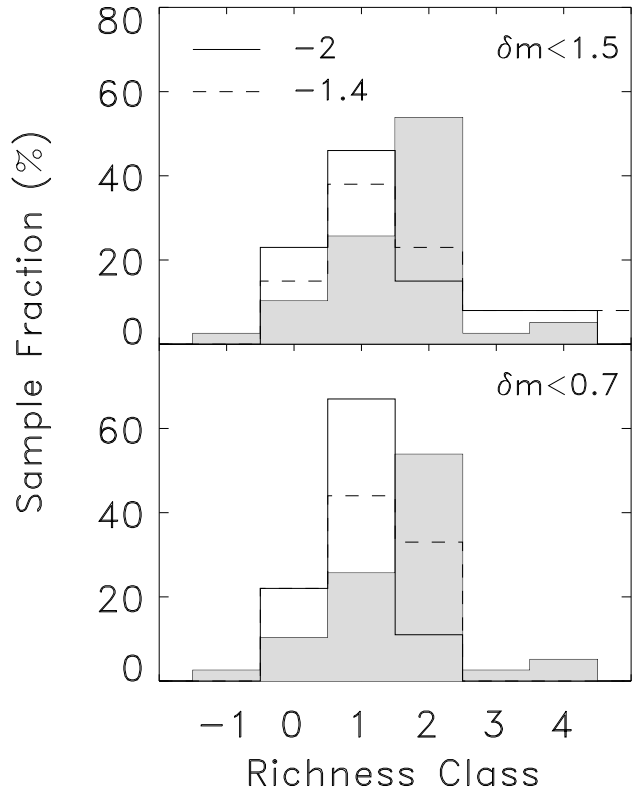}
\caption{A comparison between the distribution of richness classes for
our distant sample and the nearby X-ray sample of
Edge et al. (1990) [shaded region]. The solid histogram shows the
distribution of clusters corrected by an $r^{-2}$ surface density
profile, while the broken distribution has been corrected
using a $r^{-1.4}$ profile. The differences between the distant
and nearby samples are statistically marginal.
 \label{fig1}}
\end{figure}

We compare our distant cluster richnesses
to the nearby $z<0.18$, X-ray-flux-limited sample of
Edge et al. (1990), shown as the
filled histogram in Figure 6.  The richness classes and counts
for most clusters in the Edge et al. sample were taken from the
ACO catalog. Others were taken
from Bahcall (1980) and Owen et al. (1997).  Only clusters with
X-ray luminosities exceeding our lower limit of $L_{\rm X}(0.5-2)\sim
0.5\times 10^{44}~ \ergsec$ are included.  Edge's X-ray luminosities
were adjusted downward by typically a factor of two to match
the $ROSAT$ band. Richnesses for five Edge et al. clusters (3C129, 0745--191, A3158, Ophiucus,
Triangulum Australis) are unavailable in the literature
and were thus excluded.

Overall, the richness distributions for the 160 degree survey
clusters are similar
to the nearby cluster distribution, if not 
somewhat poorer (the modal richness of the Edge et al. clusters 
is $RC\sim 2$, while for the 160 degree survey clusters it is $RC=1$).
This difference turns, however, on only
a few clusters. Our sample is simply too small to reliably discriminate
between these distributions.   We conclude, as we did
with Figure 4, that to within the sampling precision,
the nearby and distant clusters have similar richness distributions.

\subsection{The X-ray Luminosity--Richness Relation}

In Figure 7, we plot the 14 clusters in our sample 
on the  $L_{\rm x}$--richness 
plane.  For comparison, we plotted the mean $L_{\rm x}$--richness 
relation for $<z>=0.17$ clusters from
the $ROSAT$ All Sky Survey (Briel \& Henry 1993).
Our clusters are plotted as solid points, while the
Briel-Henry relation is the solid line.  In addition,
we plotted optically-selected, $z>0.5$ clusters
from the Palomar Distant Cluster Survey (PDCS; Postman et al. 1996,
Holden et al. 1997) as open points. 
The optical cluster X-ray data from Holden et al. 1997
and the Briel \& Henry relation
have been adjusted slightly to register the X-ray passbands.
The horizontal broken line at 
$L_{\rm x}=0.5\times 10^{44}~\ergsec$ shows the approximate
luminosity of a $z= 0.5$ cluster corresponding to our X-ray flux 
detection threshold  of $f_{\rm x} \sim 2\times 10^{-14}~\ergsec~{\rm cm}^{-2}$.

Figure 7 shows the larger fraction of our clusters lying
above the Briel \& Henry mean relation in such a way
as to appear poorer than
average, as was found in Figure 6.  Nevertheless,  our clusters lie well 
within the large scatter of the Briel \& Henry relation.
Given the combination of the small sample size, the large uncertainties
in the richness corrections, and the ample spread in the
Briel--Henry relation, we can only conclude
that our clusters may be a bit poorer than average,
but they are within the expected range for nearby
clusters with similar X-ray luminosities.

\begin{figure}
\plotone{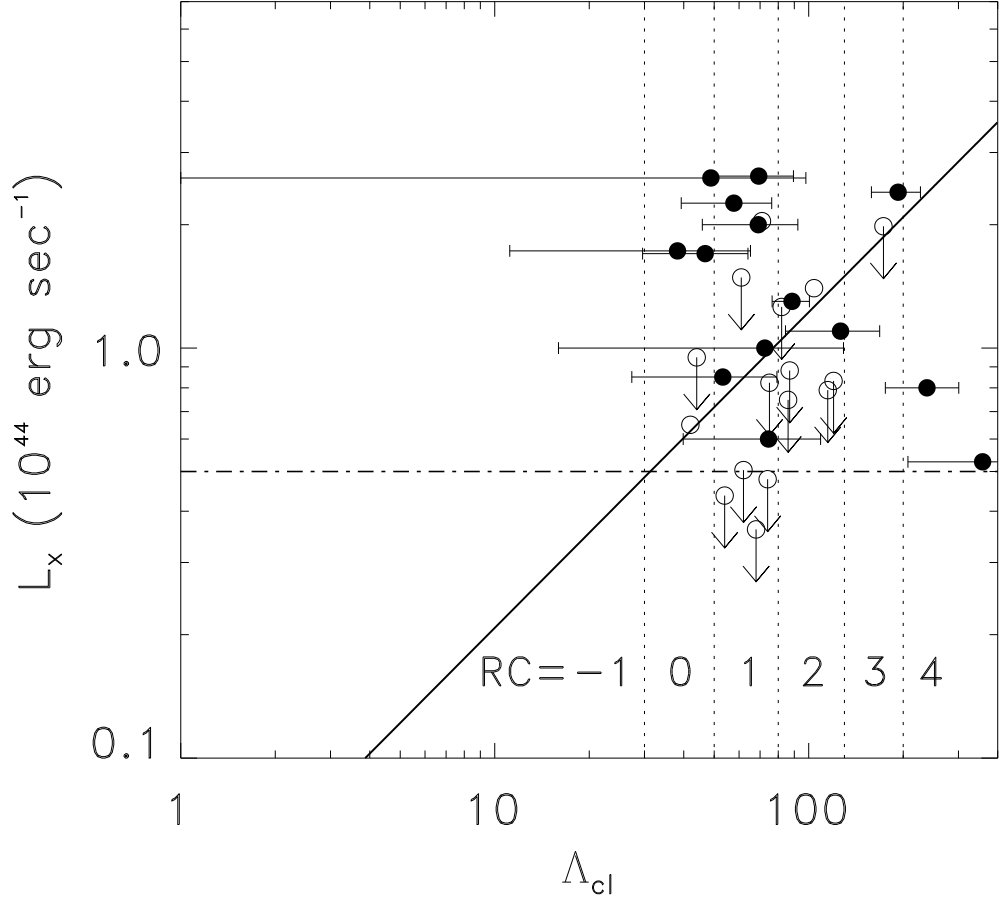}
\caption{The $L_{\rm x}$--Richness relation for our sample (solid points),
$z\gae 0.5$ PDCS clusters (open points), and the local relation 
from Briel \& Henry (1993) [solid line].  The dashed-dotted line
shows the flux limit of our survey. The richness estimates for 
the three richness class
3 and 4 clusters depend on large richness corrections.  Therefore, they may
have been systematically overestimated.  \label{fig1}}
\end{figure}

On the other hand, the optically-selected, PDCS clusters
have a similar richness distribution to our X-ray clusters, 
yet few, if any, are detected in X-rays. 
Only 3 of 16 similarly distant PDCS clusters were 
detected in X-rays, and even these may be
background sources  (Holden et al. 1997).
A similar trend was noted by Bower et al. (1994) using clusters
selected from the Couch et al. (1991) catalog.
The distributions of our clusters and the PDCS clusters
shown in Figure 7 are clearly segregated by X-ray luminosity.
Therefore, based on the analysis of these small samples,
our survey and the PDCS are apparently not
detecting similar mass concentrations.

The reason why this would be so is not entirely clear. 
However, a serious and well-known problem associated 
with optically-detected clusters
would be the tendency to misidentify superposed but physically-distinct
groups of galaxies along a sight line as a physical
cluster (see Postman et al. 1996, Oke, Postman, \& Lubin 1998).
Furthermore, optical catalogs may be prone to  contain
real, galaxy-rich mass concentrations which have not yet
virialized (Peebles 1993), yet they can be confused with rich,
virialized clusters.  On the other hand, optical selection techniques
would be sensitive to optically rich but X-ray-faint galaxy clusters
that would otherwise be missed by X-ray surveys.
This last hypothesis, intriguing as it may be, is  difficult
to understand, as significant X-ray emission is expected from clusters
in most cosmological scenarios (Bower et al 1997).

In a recent study of the Couch et al. (1991) catalog of 
clusters, Bower et al. (1997) found
concordant galaxy velocities for most clusters.
Based on an analysis of the radial velocity histograms, they 
argued that the Couch objects were real mass concentrations.
However, their velocity dispersions were nearly a factor of two larger
than nearby clusters with comparable 
X-ray luminosities.  Conversely, their X-ray luminosities were an
order of magnitude smaller than nearby clusters with similar velocity
dispersions.  Bower et al. (1997) concluded that the 
Couch cluster velocity dispersions do not reflect the clusters' virial
temperature.  Rather, the velocity dispersions are either inflated
by infalling galaxies on nearly radial orbits, or the
clusters are embedded in large-scale filaments oriented along
the line of sight.  Since the infall scenario
seems to imply an unrealistically rapid rate of infall, the
latter scenario seems more likely to be true.  The essential point
is that optical clusters may poorly represent the general population
of distant clusters (cf. Donahue et al. 2001). The comparison between our
clusters and the PDCS clusters, shown in Figure 7, supports
this interpretation.  Most importantly, the new results in this
paper demonstrate that X-ray emission is a reliable tracer of
cluster-scale mass concentrations in the distant Universe.

\section{Conclusions}

We have estimated richnesses for some of the most distant clusters
from the 160 square degree cluster survey (Vikhlinin et al. 1998a, b)
using three methods.  We found that our survey is efficient 
at detecting rich galaxy contrations at redshifts beyond $z=0.5$.
These clusters have X-ray luminosities 
between $L_{\rm x}(0.5-2)\sim 0.5-2.6\times 10^{44}~ \ergsec$,
and Abell richness classes typically between $0-2$.
The richness distribution of
the distant, 160 square degree clusters is similar to the distribution
of nearby clusters
with similar X-ray luminosities.  Clusters have
evidently evolved by less than $0.5-1$ richness classes
between redshifts of $1/2$ and today.  We find marginal evidence
that our distant clusters are somewhat poorer than the average nearby
cluster, although deep, multicolor imaging would be required to confirm
this.  

We compared our sample richnesses to those for a sample of equally distant,
optically-selected clusters 
from the Palomar Distant Cluster Survey that have been observed
by $ROSAT$ (Holden et al. 1997).  Although the PDCS clusters
are comparably rich,  they have
significantly lower X-ray luminosities compared to our clusters.
This segregation in $L_{\rm x}-$richness plane
is not understood, but may reflect the tendency for optical
surveys to detect superpositions of galaxies, and 
unvirialized mass concentrations associated with
large-scale filaments.

\acknowledgments B.R.M. thanks Larry David, Craig Sarazin, Marijn
Franx, and Tony Tyson for helpful discussions, and acknowledges support
from STScI grant GO--07533.01--96a and NAS8--39073.  We thank the
referee for suggestions that substantially improved this paper.

\clearpage

\begin{table}
\caption{Distant X-ray Clusters }
\vspace{2.5 mm}
\begin{center}
\baselineskip=0.01mm
\begin{tabular}{lcccccccc}\hline\hline
Cluster &$z$& $f_{\rm x}(0.5-2)$&$L_{\rm x}(0.5-2)$&$\delta m$& $\Delta m$ &$\sigma_{\rm cl}$&$n_1$ &RC\\
\hline 
~~&~~&$10^{-14}~{\rm cgs}$&$10^{44}~{\rm erg~s}^{-1}$&mag&mag&~~~& ~~~& ~~~\\
\hline
0030$+$2618 & 0.500&  24.3 & 2.6   &0.43 &2.12 &2.1 &$20\pm 11$ &1\\
0521$-$2530 & 0.581&  17.6 & 2.6   &0.79 &1.72 &... &$-5\pm 9$  &0\\
0522$-$3625 & 0.472&  18.4 & 2.3   &0.03 &2.51 & 2.0&$56\pm 13$ &1\\
0529$-$5848 & 0.466&  ~~5.6& 0.6 &0.00 &3.0 & 1.6&$64\pm 15 $ &1\\
0847$+$3449 & 0.560&  12.2 & 1.7   &0.55 &1.59&  1.4&$18\pm 10$ &0\\
0848$+$4456 & 0.574&  ~~3.3 & 0.5   &1.73 & 0.57& 2.8&$22\pm 12$ &4\\
0858$+$1357 & 0.500 & ~~6.4  & 0.8  &0.99 & 1.59& 2.1&$21\pm 9 $ &4\\
0951$-$0128 & 0.567 & ~~ 7.1 & 1.0 &0.58& 1.99& 2.9&$14\pm 8 $ &1\\
0956$+$4107 & 0.600  & 15.6 &2.4    &1.44 & 0.92& 1.2&$15\pm 8 $ &3\\
1002$-$0808 & 0.520 &  ~~8.6 & 1.1 &0.96& 1.42& 2.3&$27\pm 9 $ &2\\
1311$-$0551 & 0.461 &  13.7 &1.3   &0.00 & 3.0& 2.2&$57\pm 15 $ &2\\
1354$-$0221 & 0.546  &14.5 & 2.0   &0.00 & 2.91& 3.0&$47\pm 11$ &1\\
2146$+$0422 & 0.531& 13.8 & 1.7     &0.21 & 2.07& 1.0&$5\pm  6 $ &0\\
2328$+$1453 & 0.497 & ~~7.6 & 0.9  &0.08& 2.51& 2.2&$35\pm 11$ &1\\
\hline\hline
\end{tabular}
\end{center}
\end{table}

%\clearpage

\end{document}